\documentclass[aps,prl,twocolumn,showpacs]{revtex4}
\usepackage{graphicx}
\usepackage{dcolumn}
\usepackage{amsmath}
\usepackage[latin1]{inputenc}

\begin{document}

\title[Extracting the Landscape and Morphology  of Aging Glassy Systems]{ Extracting the Landscape and Morphology  of Aging Glassy Systems}
\author{Paolo Sibani}\email[Corresponding author ]{paolo@planck.fys.sdu.dk}
\author{Jesper Dall}
\affiliation{Fysisk Institut, Syddansk Universitet, Campusvej 55, 
DK--5230 Odense M, Denmark} 

\date{\today} 

\begin{abstract}
{\small  
We propose a way to analyze
 the landscape geometry  explored  by a glassy system  
 after a quench solely based on  time series of  energy values   
recorded  during a    simulation.  Entry and exit times
 for landscape `valleys'
 are  defined operationally   by  the occurence
of  anomalous  energy changes 
revealed  by barrier and energy records.
Linking these    non-equilibrium events---or `earthquakes'---
to  the  record statistics of the thermal noise   immediately   leads to 
the $\ln t$  relaxation behavior  ubiquitous in glassy dynamics. 
Aging of Ising spin glasses in two 
and three dimensions is  studied as a check    for 
a   number of low temperatures and lattice sizes.
A  simple  picture emerges on e.g. 1) the 
scaling  with temperature and system size of
 the energy barriers as a function of   $\ln t$,
 2) the scaling with system size of the  
lowest energy  seen in  each valley.   
} 
\end{abstract}

\maketitle


Macroscopic   properties  
of  thermalizing glassy systems   depend on the time (age)    
elapsed after the quench into the glass
phase:      on  time scales  shorter  than 
 the age, a state of  pseudo thermal equilibrium  
  holds  locally  within  metastable
regions of the landscape, while on  longer observational time scales
the   relaxation is  manifestly non-stationary~\cite{Alba86,Svedlindh87,Andersson92}.
Some models  of  aging  build on heuristic  assumptions 
on the  morphology of
low energy excitations in real space~\cite{Fisher88a,Koper88},
while others  start from a coarse grained
state space~\cite{Sibani89,Sibani91,Joh96,Hoffmann97} or, more 
simply, assume a given distribution of exit times from
traps in configuration space~\cite{Bouchaud92}.  
Still, two related and  fundamental 
questions remain unanswered:
 1) which dynamical events mark the transition between 
equilibrium and non-equilibrium dynamics, and 
2) how can the  corresponding, \emph{dynamically selected}, structures 
in configuration and/or  real space  be identified?

In  models of  driven dissipative systems with
multiple attractors~\cite{Coppersmith87,Sibani93a,Sibani01},
marginally metastable attractors with an \emph{a priori}
 negligible  statistical weight  
are nevertheless  typically selected  by the dynamics,
an effect which  underlies   memory and rejuvenation effects
  analogous to  those observed in the thermalization
of e.g.\ spin glasses~\cite{Jonason98} after a quench.
One can   speculate that similar mechanisms  could  generally  be  present 
in  glassy systems with an extensive
number of metastable attractors. To   guarantee the dynamical
relevance of the attractors  found  and/or their real space counterparts,   
the algorithms   utilized should match   the   dynamics in 
question, e.g.   thermalization after a deep  quench. 
This however is not the case for  standard procedures  as e.g. the Stillinger-Weber
approach~\cite{Stillinger83,Nemoto88,Becker97,Crisanti02,Mossa02},
genetic algorithms~\cite{Palassini99}
and   energy minimization of excitations of  fixed volume~\cite{Houdayer00,Lamarcq02}. 
  
A complementary  approach is  presented below
which   solely   relies on  statistical  information   
collected   during the unperturbed relaxation
following an  initial quench. The approach   identifies
a temporal sequence of   non-equilibrium events, or
`earthquakes'. These    entail   major 
reorganizations of the landscape (1) and
subdivide the  states    visited  
into an ordered   sequence of  sets, or   `valleys'
wherein (2)   the equilibrium-like fluctuation 
dynamics unfolds. 
  Both (1) and (2) 
are required for a consistent  description,  
and both  properties have  been  checked extensively  for
short range spin glass models in 2d and 3d.
For these models, the non-equilibrium aging  
of physical  quantities as e.g. the size of the barriers
overcome   depends simply   
on the  `valley index'. This  is a proxy for 
the system age   marking  the position
of each valley within the ordered sequence
 of valleys visited.

\begin{figure}[ht]
\begin{center}
\includegraphics[width=8cm]{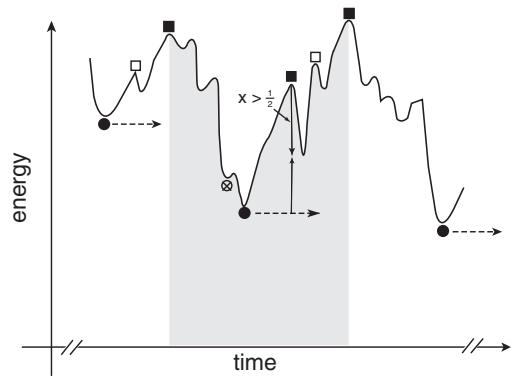}
\vspace{-0.2cm}
 \caption{\small A fictitious time series
of collected energy values   
illustrates  our  sampling
procedure and our  definition
 of a valley (= gray area). 
   Conspicuous events, i.e.\ 
energy and barrier records, 
 are denoted by circles and squares. }
 \label{fig1}
\end{center}
\vspace{-0.6cm}
\end{figure} 

A  discussion of the  
free energies associated with the  fluctuation
regime inside the valleys is left   
to    forthcoming publications and  
the variable $t$ will  
denote the age of the system, except
when otherwise indicated.

{\bf Method}: A multitude of  inequivalent 
metastable states  produces  the 
time inhomogeneous dynamics
generally observed in  aging    systems. 
In most  cases,  these states are sets
of configurations surrounding 
 local energy minima, i.e.,
 qualitatively speaking, valleys in the energy landscape.   
Since    high-lying and
shallow local energy minima   
greatly outnumber deep minima,  
the initial quench   typically produces  a  poor local minimum and
  an   aging  system subsequently   encounters   a series of 
progressively higher energy
barriers~\cite{Sibani98,Joh99,Lederman91}
which delimit  progressively deeper valleys.
The values of the  highest barrier $B$  and lowest energy 
$E$---henceforth \emph{best so far} or simply bsf 
barriers and energies---stand out
as crucial events, and 
  will serve to  tag  the valleys  as they  appear 
in the energy landscape.

The simulations start  from a  random, $T=\infty$, 
configuration  and run  at   temperature $T$. 
Barrier values are  computed as energy differences from
the current bsf energy state. 
The bsf energy and barrier are  initially   $\infty$  and 
 zero, respectively.\ 
Each run produces  a time ordered
sequence  of $E$'s and $B$'s as  e.g.\  
$EBBEEBBBE$ in Fig.~\ref{fig1}.  
Contiguous bsf energies 
pertain to a trajectory   `tumbling'
down  in a valley. Being   mainly interested
 in the configuration  where the tumbling stops,
we only keep  the lowest $E$  within each   group.
 In Fig.~\ref{fig1} the  
datum  removed is  represented by the  `$\otimes$' symbol. 
 Similarly, we want to   avoid registering   the numerous   bsf 
 barriers of roughly the same height which  appear in close succession
while  the system 
goes through a high `saddle'.
 A barrier record is therefore kept
only if the  energy    experiences a 
sufficiently deep local minimum   before
the next  global barrier maximum is encountered. 
The precise definition of `sufficiently deep' involves 
a positive number  $x \leq 1$. For $x=1$,   the
local  minimum  must  be as low 
  as  the current bsf energy, while
for $x=0$ all bsf barriers are kept.  
Interestingly, our results are  insensitive to the value of $x$
in the range $0.25-0.75$. 
In terms of Fig.~\ref{fig1}, we only keep
the full squares and circles, thereby reducing our string 
even further to its final state: $EBEBBE$.  

For   speed,  we rely on the Waiting Time Method
(WTM), a    rejection-less Monte Carlo scheme~\cite{Dall01}    
well suited for problems where $N$ variables
contribute additively to the energy   through
 local interactions. 
Based on  the local field,
each variable is stochastically assigned a  
\emph{flipping  time} and the variable 
with the shortest flipping time is  
updated together with the  local 
fields affected by the move.
The   sequence of WTM moves   
 equals  in probability~\cite{Dall01}  
 the  sequence of accepted moves
in the Metropolis 
algorithm. 
The current flipping   time, henceforth
simply   `time',  corresponds to   a  Metropolis sweep,
and to    the  physical  time of an experiment. 
 
 {\bf Results}: We consider   $L^d$ Ising spins 
placed on a regular lattice
of linear size $L$ 
which interact via   the Edwards-Anderson 
Hamiltonian~\cite{Edwards75}
\begin{equation}
{\cal H}(\alpha)   = -
\frac{1}{2}\sum_{i,j} J_{ij}s_i^{(\alpha)}s_j^{(\alpha)}.
\end{equation} 
In this formula,  $s_i^{(\alpha)}$
 is the spin value at site $i$ 
for   configuration $\alpha$, 
the couplings     $J_{ij}$ are
symmetric in their two indices and  vanish 
unless $i$ and $j$ belong 
to neighbor sites. In  the
latter case they are 
drawn independently 
from a Gaussian distribution of unit variance.
The dynamics 
is  nearest neighbor thermal hopping with  
detailed balance. 
The runs spanned over
several time decades, always bringing the 
 system down to quite low  energies, typically 
 around $-1.66$ per spin.   
 Large systems tend to have more shallow valleys
 than smaller system. To obtain meaningful 
 comparisons   all data originating from the 
 first ten  time units after the quench were discarded,
 and   the   value  of the valley index was 
 shifted up to one
 unit in our scaling plots in Figs.~\ref{scaling_B}~and~\ref{fig4}. 

\begin{figure}[t]
\begin{center}
\includegraphics[width=7.5cm]{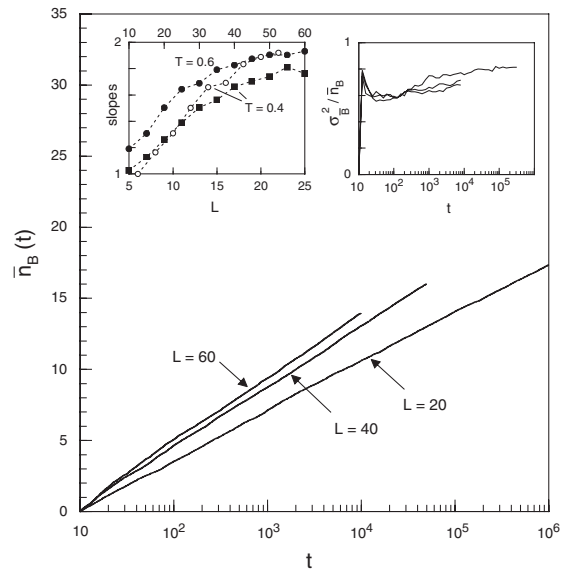}
\vspace{-0.2cm}
\caption{\small Main figure: the average number $\overline{n_B}$ 
of barrier records versus time for three selected system sizes in $2d$
at $T=0.4$.
Left insert: the logarithmic slopes of $\overline{n_B}$ as a function
of the linear system size $L$  for $2d$ ($ T=0.4 $ and $0.6$) and $3d$ ($T=0.4$).
The upper (lower) scale refers to $2d$ ($3d$) systems.
Right insert: the variance to  average ratio for $n_B(t)$  
for the  data in the main figure. }
 \label{B_in_time}
\end{center}
\vspace{-0.5cm}
\end{figure}  

A basic quantity is the number of events
(barrier or energy records) which on average occur in 
$[0,t]$. Since achieving a record becomes
increasingly difficult, the rate of events will 
 decrease in time and the dynamics
decelerates.  Interestingly, a time homogeneous 
 dynamics is restored by  using the natural logarithm
 $\ln t$ as   time variable. This is  shown in 
Fig.~\ref{B_in_time}, where  
the average number $\overline{n_B}(t)$ of barrier crossing events 
is plotted as a function
 of   $\ln t$.  The data shown in the main figure  belong  to 
 $2d$ systems, while similar data in both $2d$ and $3d$ 
  were  omitted for clarity. The left insert  
shows the logarithmic rate of events $d \overline{n_B}(t)/ d \ln t$, 
in $2d$ as well as $3d$, as a function of 
$L$.
Finally, the right insert shows  the 
ratio of the  variance $\sigma^2_{n_B}(t)$ 
to   $\overline{n_B}(t)$. Both  
quantities were  estimated for   $3000$  independent runs:  
The variance and the average follow  the same logarithmic
fashion, with the former trailing the latter due to 
an initial lag. A similar statistics is also approached by  the 
number of bsf records---and hence 
the number of valleys---discovered on average in  $[0,t]$, 
for sufficiently large systems.
A logarithmic dependence of macroscopic averages is ubiquitous
in complex metastable  systems. Beside the  present examples
 it is found in e.g. driven dissipative models~\cite{Sibani01}, 
 vibrated granular systems~\cite{Jaeger89,Josserand00}, 
 Lennard-Jones glasses~\cite{Kob00}, nitridic ceramics~\cite{Hannemann02}
and population dynamics~\cite{Sibani99}.
  
The approximate equality 
between   average and variance points to 
the relevance of   \emph{log-Poisson}
statistics  describing  the number
of magnitude records  achieved in  a sequence of $t$ 
independent random numbers~\cite{Sibani93a}.
The name   stresses    that 
$\ln t$, rather than  $t$, is the relevant
time variable. 
For data   stemming  from 
a nearest neighbor random walk in  a  \emph{landscape}, 
log-Poisson statistics  indicates  
 a \emph{de-correlation} of the   energy function     
 between consecutive  
barrier records, and \emph{a fortiori}
 during the time spent within each  valley.
Since   the  correlation time and  the relaxation time
(for  the internal thermalization) can be 
generically identified~\cite{VanKampen92},  
 we   conclude that      thermal quasi-equilibrium  
is   achieved within a  valley. As a consequence, 
the morphology of  quasi-equilibrium fluctuations around
the  bsf state of  each  valley can be
isolated from the non-equilibrium dynamics
and analyzed  separately. 
  
\begin{figure}[t]
\begin{center}
\includegraphics[width=7.5cm]{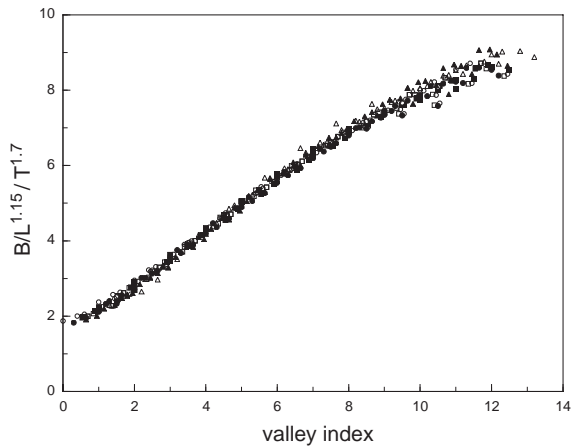}
\vspace{-0.2cm}
\caption{\small Average energy barriers records  $B$ separating contiguous valleys in $3d$
 spin glass landscapes are  scaled with  $L$ and  $T$, as indicated in the ordinate label
 and plotted versus $n$.
All combinations of 
 $L=8,10,12,14,16$ and $T=0.3, 0.4, 0.5, 0.6$ and $0.7$
 are included, plus  $L=18$ and  $T=0.4$.}
 \label{scaling_B}
\end{center}
\vspace{-0.5cm}
\end{figure} 

\begin{figure}[t]
\begin{center}
\includegraphics[width=7.5cm]{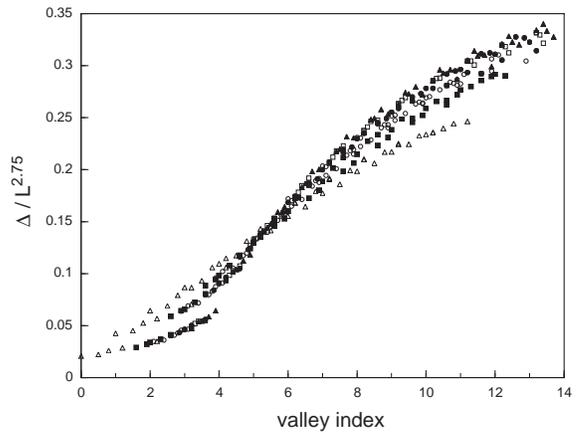}
\vspace{-0.2cm}
\caption{\small The  average
difference $\Delta$ between the lowest  energies  of  the first
and the  $n$'th  valley 
 is plotted versus $n$. The ordinate  is scaled with   $L$  
as indicated.  The  system and the combinations of $L$ and $T$ shown are 
as in Fig.~3, except for an additional 
 $T=0.2$ data set (upper triangle), which  
  falls outside  the main trend.  The remaining spread in the  data collapse 
cannot be removed by introducing an additional scaling parameter, 
and is likely indicative of an underlying distribution
of minima.   
}
 \label{fig4}
\end{center}
\vspace{-0.5cm}
\end{figure} 

The scaling of the
logarithmic rate of events with $L$ is   
 shown in    the left
insert of Fig.~\ref{B_in_time}
to increase   from
an initial value of $1$ to approximately 
$2$ for large system sizes. 
With   Poisson statistics,   
 this rate is  proportional  to
  the number of  lattice  `regions' evolving independently. 
  As the spins are quiescent most of the time, 
 a record in  one   region   leads
to an overall record as well. 
Thus,   two   length 
scales~\cite{Lamarcq02} might in principle  coexist:
the  linear size
 of the quasi-equilibrium 
fluctuations  and  
  the  extent 
of the non-equilibrium `earthquakes'   
leading  from  one valley to the next. 
Figs.~\ref{B_in_time} and~\ref{fig4} imply  that  
the latter scale is  asymptotically
$\propto L$, i.e. earthquakes are
extensive or near extensive events. This is further  confirmed by  
 the distribution (not shown) of the
Hamming distance (i.e. the  
  number 
of spins whose  orientations differ), 
between the bsf configurations of 
 valleys $n$ and $n+1$.
 The distribution is  exponential,
and its   average   
grows   linearly  in   
$n$ and close to linearly in  $L^d$.   
 
The scaling with $L$ and $T$ 
 of the energy barrier $B$ separating two
contiguous valleys 
 in $3d$ systems  is 
shown in Fig.~\ref{scaling_B}.
All points are   quenched averages 
 over $3000$ independent runs 
 for the $L$ and $T$
 values specified  in the caption.
 As mentioned, the data have been shifted by up to 
 one unit along the abscissa, which mirrors 
 the arbitrariness of the initial 
 bsf states---conventionally
    recorded  after ten time units.    
As indicated in the plot, 
$B \propto L^{1.15} T^{1.7}$. We have not attempted
a quantitative determination of the uncertainty 
on the scaling exponents, but changing the last significant 
digit visibly affects the quality of the plot. 

The above considerations  apply equally well to Fig.~\ref{fig4},
which, for the same set of 
runs,  depicts the average difference $\Delta(1,n)$  between the bsf
energy of valley $1$ and $n$.  $\Delta(1,n)$ is independent
of $T$, as one should expect, and increases almost linearly 
with the valley index. Refs.~\cite{Sibani91,Hoffmann97} analyze 
 a mesoscopic model with this exact property.   
The curvature for large $n$ might partly be due to
a systematic error, since not every run enters the last few valleys
within the maximum time span allotted to the calculations.
For very long time scales, the curve will approach the 
constant value of the 
difference between the first bsf energy and the energy of the
ground state.  
  
The    strong $L^{2.75}$    size dependence of $\Delta(1,n)$ is
 noticeable: since  moving from one valley to the other
 involves the rearrangements 
of ${\cal O} (L^3)$ spins, it seems likely  that  the
 set difference between two bsf energy configurations 
be a sponge-like object of the sort 
discussed in~\cite{Houdayer00,Lamarcq02}.
The energy difference between different 
bsf states is clearly  not ${\cal O}(1)$ as 
expected in  mean-field. It   
 seems unlikely that going to lower energies
could  radically change the scaling of $\Delta$ with system
size. 
We stress  again
that    the WTM  and the Metropolis
algorithm being   equivalent,    
everything could  have been 
done  in the  conventional way,  
albeit with  much greater  numerical effort. 

In summary,   the
distinction between pseudo-equilibrium fluctuations and
strongly non-equilibrium events in aging dynamics is
emphasized. We  link the latter events to barrier and energy records 
 marking the appearance of    new  valleys 
 as the   landscape  unfolds during aging.   
The  homogeneity of the event statistics 
in  $\ln t$ is explained  theoretically  
in terms of the    statistics of records.

The description of spin glass non-equilibrium
dynamics as a   log-Poisson process indicates that
thermal aging can be treated on the same footing as  
 driven dissipative dynamics~\cite{Sibani93a,Sibani01} and
evolution models~\cite{Sibani99a,Hall02}, in spite 
of  obvious differences at the  microscopic level.
It also  implies  that, on each
time scale,  glassy dynamics typically selects the  least 
stable   among the metastable configurations  available. 
While this  feature has    been noticed in more restricted
contexts~\cite{Coppersmith87,Sibani93a,Sibani01}, the 
present study suggests that it might provide 
a powerful and general paradigm for  glassy
relaxation.

\noindent {\bf Acknowledgments}:
PS is indebted to  Greg Kenning and Henrik J. Jensen for discussions
and to the Danish SNF for grant 23026.
\bibliographystyle{unsrt}

\bibliography{SD-meld,thesis}

\end{document}